\documentclass[prd,twocolumn,superscriptaddress,nofootinbib,amsmath,amssymb,preprintnumbers,floatfix]{revtex4-2}

\usepackage{graphicx,color,amsmath,amssymb}
\usepackage{slashed}
\usepackage{braket}
\usepackage{lipsum}
\usepackage{graphicx}
\usepackage{aas_macros}
\usepackage{physics}

\newcommand{\be}{\begin{equation}}
\newcommand{\ee}{\end{equation}}
\newcommand{\nocontentsline}[3]{}
\newcommand{\toclesslab}[3]{\bgroup\let\addcontentsline=\nocontentsline#1{#2\label{#3}}\egroup}
\newcommand{\tocless}[2]{\bgroup\let\addcontentsline=\nocontentsline#1{#2}\egroup}

\newcommand{\azero}{{\rm a}_0}

\newcommand{\Htwo}{\rm{H}_2}

\usepackage[export]{adjustbox}
\usepackage{tikz}
\usepackage{tkz-euclide}
\usetikzlibrary{decorations.pathmorphing}	
\tikzset{
    v/.style={decorate, decoration={snake, segment length=3mm, amplitude=0.75mm}, draw},
    f/.style={draw=black, postaction={decorate},
        decoration={markings,mark=at position .6 with {\arrow[very thick]{latex}}}},
    fb/.style={draw=black, postaction={decorate},
        decoration={markings,mark=at position .4 with {\arrowreversed[very thick]{latex}}}},
    fnar/.style={draw=black},
    g/.style={decorate, draw=black,
        decoration={coil,amplitude=3pt, segment length=3.5pt}},
    s/.style={dashed,draw=black, postaction={decorate},
        decoration={markings,mark=at position .55 with {\arrow[very thick]{latex}}}},
    sb/.style={dashed,draw=black, postaction={decorate},
        decoration={markings,mark=at position .55 with {\arrowreversed[draw=black,very thick]{latex}}}},
    snar/.style={dashed,draw=black,line width =1.25pt},
}

\usepackage{hyperref} 
\usepackage{amsmath}
\hypersetup{
    colorlinks=true,
    citecolor=blue,
    linkcolor=magenta,
    filecolor=magenta,
    urlcolor=blue,
}

\newcommand{\keV}{{\, {\rm keV}}}

\newcommand{\LL}{{\mathcal{L}}}

\newcommand{\htwo}{H$_2$ }
\newcommand{\zetah}{$\zeta^{\text{H}_2}$}
\newcommand{\htp}{{\rm H}_3^+}
\newcommand{\zetacr}{\zeta^{{\rm H}_2}}

\definecolor{mypurple}{RGB}{164,64,214}

\begin{document}

\title{Constraining Dark Matter-Proton Scattering from Molecular Cloud Ionization}



\author{Carlos Blanco}
\email{carlosblanco2718@princeton.edu}
\affiliation{Department of Physics, Princeton University, Princeton, NJ 08544, USA}
\affiliation{Stockholm University and The Oskar Klein Centre for Cosmoparticle Physics,  Alba Nova, 10691 Stockholm, Sweden}

\author{Ian Harris}
\email{ianwh2@illinois.edu}
\affiliation{Department of Physics, University of Illinois Urbana-Champaign, Urbana, Illinois 61801, U.S.A.}
\affiliation{Illinois Center for Advanced Studies of the Universe, University of Illinois Urbana-Champaign, Urbana, Illinois 61801, U.S.A.}

\author{Yonatan Kahn}
\email{yfkahn@illinois.edu}
\affiliation{Department of Physics, University of Illinois Urbana-Champaign, Urbana, Illinois 61801, U.S.A.}
\affiliation{Illinois Center for Advanced Studies of the Universe, University of Illinois Urbana-Champaign, Urbana, Illinois 61801, U.S.A.}

\author{Anirudh Prabhu}
\email{prabhu@princeton.edu}
\affiliation{Princeton Center for Theoretical Science, Princeton University, Princeton, NJ 08544, USA}

\date{\today}

\begin{abstract}

Optically dense clouds in the interstellar medium composed predominantly of molecular hydrogen, known as molecular clouds, are sensitive to energy injection in the form of photon absorption, cosmic-ray scattering, and dark matter (DM) scattering. The ionization rates in dense molecular clouds are heavily constrained by observations of abundances of various molecular tracers. Recent studies have set constraints on the DM-electron scattering cross section using measurements of ionization rates in dense molecular clouds. Here we calculate the analogous bounds on the DM-proton cross section using the molecular Migdal effect, recently adapted from the neutron scattering literature to the DM context. These bounds may be the strongest limits on a strongly-coupled DM subfraction, and represent the first application of the Migdal effect to astrophysical systems.

\end{abstract}

\maketitle

{\hypersetup{linkcolor=blue}
}

\section{Introduction}

The evidence for dark matter (DM) is overwhelming, but the absence of direct evidence for Weakly Interacting Massive Particle (WIMP) DM with masses at or above the GeV scale has motivated lighter sub-GeV DM candidates, as well as models with different DM particles as subcomponents of the total DM density. At these low masses, DM with with relatively strong coupling to Standard Model (SM) matter has distinctive phenomenology. Terrestrial laboratory searches for DM require the interaction strength of DM with baryonic matter to be sufficiently weak to permit the DM to traverse both the atmosphere and the experimental shielding. By contrast, some astrophysical targets are naturally unshielded and can directly constrain models where the DM-SM coupling exceeds thresholds for detectability in surface and subterranean detectors.



The DM-baryon interaction cross section can be indirectly, yet strongly, constrained using cosmological and astrophysical observables. In the early universe, strongly-coupled DM can distort the spectrum of cosmic microwave background (CMB) and acts to smooth out its measured temperature anisotropies~\cite{Ali-Haimoud:2015pwa, Ali-Haimoud2021,Dvorkin:2013cea, Gluscevic:2017ywp, Boddy:2018kfv, Xu:2018efh, Slatyer:2018aqg, Boddy:2018wzy}. At late times DM will also impede the formation of small-scale structure which is measurable through observations of, e.g., Milky Way satellites~\cite{Nadler:2019zrb, Nguyen2021} and the Lyman-$\alpha$ forest \cite{Viel:2013apy}. Furthermore, DM-baryonic scattering can heat up gas-rich dwarf galaxies with measurably small radiative cooling rates~\cite{Wadekar2,Wadekar2021}. A general caveat to indirect constraints from the CMB, dwarf galaxy heating, and Milky Way satellites is that a sufficiently small, strongly-coupled, DM~\textit{sub-component} becomes effectively indistinguishable from a small perturbation to the SM baryonic content. Precise determination of the DM subfraction, $f_\chi$, at which these bounds no longer apply or are significantly relaxed generically requires detailed numerical simulations, but in the case of the CMB, the bounds disappear if $f_\chi \lesssim 0.4\%$~\cite{Boddy:2018wzy}.

In a recent work~\cite{Blanco:2022mci}, a new method was introduced to constrain models with a DM subcomponent strongly coupled to electrons by studying  DM-induced ionization in molecular clouds (MCs). Cold and dense MCs are very efficient attenuators of visible and ultraviolet radiation from stars, which makes cosmic rays (CRs) the dominant source of ionization in the interior and cores of these clouds. Infrared measurements of the chemical tracers of ionization have inferred tiny ionization rates, \zetah, and free electron abundances, $x_e \equiv n_e/n_{\text{H}_2}$ (see, e.g., \cite{Draine2011}). These constraints were used to set limits on the DM-electron cross section since DM-induced electron recoils can directly ionize $\Htwo$, which for sufficiently large cross sections would exceed the observed ionization rates in MCs.

In this paper, we build on the technique of Ref.~\cite{Blanco:2022mci} by extending the analysis of MC ionization to scenarios with DM-baryonic couplings. We make use of the formalism of Ref.~\cite{Blanco:2022pkt} -- which derived the molecular Migdal effect for DM following earlier work in the neutron scattering literature~\cite{lovesey1982electron,colognesi2005can} -- to predict the molecular ionization rate following a nuclear recoil. Specifically, we calculate the ionization and dissociation probability of diatomic hydrogen due to DM-induced nuclear recoils and use its contribution to \zetah \ in order to constrain DM-proton cross sections that are inaccessible to current direct and indirect searches. This process is the astrophysical analogue of direct detection efforts to constrain sub-GeV dark matter using the Migdal effect, but to our knowledge, ours is the first prediction of any astrophysical Migdal-like excitation or ionization. In terrestrial detectors, Migdal ionization in isolated atoms has recently been used in liquid noble detectors to extend the sensitivity of DM-nuclear scattering to lighter DM masses~\cite{XENON:2019zpr,LZ:2023poo}, and several attempts to observe and calibrate this effect with neutron beams are in progress~\cite{Bell:2021ihi,Xu:2023wev}. Furthermore, theoretical and experimental efforts are underway to search for the visible signatures of DM-induced nuclear and electronic scattering in laboratory molecular detectors~\cite{Essig:2016crl,Blanco:2019lrf,Blanco:2021hlm,Blanco:2022cel,Blanco:2022pkt,Araujo:2022wjh,Bell:2023uvf}. We show in this paper that just as the Migdal effect can extend the sensitivity of DM-nuclear scattering experiments to sub-GeV masses, astrophysical MC molecular Migdal ionization can provide a robust constraint on strongly-coupled subfractions of sub-GeV DM.

The paper is organized as follows. In Sec.~\ref{sec:CRionization} we describe how the CR ionization rate is measured in MCs and discuss the clouds that we use to derive constraints. In Sec.~\ref{sec:MolecularMigdal} we review the molecular Migdal effect and use it to derive the Migdal-induced ionization/dissociation rate in $\Htwo$ as a function of the DM cross section, relic abundance subfraction, and velocity distribution. In Sec.~\ref{sec:results}, we use the observed CR ionization rates to derive bounds on the DM-proton cross sections for the two limiting cases of heavy and light mediators. We conclude in Sec.~\ref{sec:discussion} with a brief discussion of our results.

\section{Cosmic Ray Ionization in Molecular Clouds} \label{sec:CRionization}

The presence of free electrons and ions strongly affects gas dynamics in the interstellar medium (ISM). Sources of local fluctuations in the free electron abundance, $x_e$, include ionizing radiation (UV and X-ray) from hot stars, as well as shock waves from supernovae and stellar outflows. A global, irreducible source of ionization comes from CRs. Early attempts to theoretically model CR ionization in clouds made up of predominantly atomic hydrogen predict a rate per atom of $6.8 \times 10^{-18} \ {\rm s}^{-1} \lesssim \zeta^{\rm H} \lesssim 10^{-15} \ {\rm s}^{-1}$~\cite{SpitzerTomasko1968}. The upper and lower bounds on $\zeta^{\rm H}$ assume different spectra for CRs with kinetic energy less than $\sim 10$ MeV. A precise determination of the ionization rate requires direct measurement of the low-energy spectra of CR electrons, protons, and other nuclei. Determination of the CR electron spectrum from measurements of the Galactic synchrotron background~\cite{Ginzburg1965, Orlando2018, PadovaniGalli2018} suffer from large uncertainties in modeling of the interstellar magnetic field. Additionally, models of the CR proton spectrum based on measurements of local gamma-ray emissivity~\cite{Casandjian2015, Orlando2018} only constrain CRs with $\sim {\rm GeV}$ or larger energies. Low-energy CRs are significantly affected by solar modulation, including solar winds and interplanetary magnetic fields. In-situ measurements of the CR spectrum down to $\approx 3$ MeV have been made by the \emph{Voyager I, II} spacecrafts~\cite{Cummings2016, Stone2019}. Still, the cross section for ionizing atomic hydrogen peaks at $\approx 50$ eV for CR electrons and $\approx 10$ keV for CR protons~\cite{Padovani2009}, requiring alternative methods for inferring the low-energy CR spectrum. 

A powerful technique for determining the CR ionization rate involves measuring emission line intensities of molecular tracers whose abundances are correlated with the free electron abundance. The choice of tracer depends critically on the gas density and chemical composition. In diffuse parts of MCs, $\htp$ is a particularly attractive candidate due to its simple and well-understood chemistry~\cite{Oka2006, Dalgarno2006, IndrioloMcCall2013}. In denser regions such as MC cores and proto-stellar clusters, tracers such as HCO$^+$, DCO$^+$, and CO are often used~\cite{Caselli1998}. In the following sections, we provide some details regarding the inference of CR ionization rates in diffuse and dense MCs using abundance measurements of various tracers.

\subsection{CR ionization rate from ${\rm H}_3^+$}

In diffuse MCs which contain a significant abundance of $\Htwo$, but not other molecular species, the chemical network used to predict the $\htp$ abundance is

\begin{align}
    \Htwo + {\rm CR} &\to {\rm H}_2^+ + e^- + {\rm CR} \label{eqn:crchem}\\
    {\rm H}_2^+ + \Htwo &\to {\rm H}_3^+ + {\rm H}  \label{eqn:h3form}\\
    {\rm H}_2^+ + e^- &\to {\rm H} + {\rm H} \label{eqn:dissrec}\\
    {\rm H}_2^+ + {\rm H} &\to \Htwo + {\rm H}^+ \label{eqn:chargetransfer},
\end{align}

\noindent where (\ref{eqn:crchem}) occurs before the other processes. The process in (\ref{eqn:dissrec}) corresponds to dissociative recombination and (\ref{eqn:chargetransfer}) to charge transfer. These processes reduce the equilibrium $\htp$ abundance. Similar processes to (\ref{eqn:crchem}) occur due to photoionization and, as we show in this paper, DM; however we neglect photoionization in our analysis as it is subdominant to CR ionization~\cite{IndrioloMcCall2013}. Ionization of $\Htwo$ is the bottleneck for the chemical network, and thus the CR ionization rate should be approximately equal to the formation rate of $\htp$. In diffuse clouds, with moderate ionization fractions, $\htp$ can absorb electrons, resulting in either $\Htwo$ and H or three hydrogen atoms. In dense regions, with lower ionization fraction, $\htp$ transfers a proton to neutral molecules in the cloud such as O, CO, and N$_2$, resulting in $\Htwo$ and OH$^+$, HCO$^+$, and HN$_2^+$, respectively.

The abundance of $\htp$ can be determined by measuring the intensities of various rovibrational transitions. In most molecular clouds, only the ground states -- corresponding to angular momentum quantum numbers $(J, K) = (1, 0)$ (ortho) and $(J, K) = (1,1)$ (para) -- are occupied.\footnote{The quantum number $J$ refers to the total angular momentum, while $K$ represents the projection of the angular momentum onto the molecular axis.} The observed infrared emission results from transitions to ro-vibrational excited states followed by de-excitation to the ground state. The column density of $\htp$, $N(\htp)$, can be determined from the widths and intensities of these transitions. In steady state, reaction (\ref{eqn:crchem}) relates the CR ionization rate to the column density of $\htp$ through~\cite{Geballe1999}

\begin{align}
    \zetacr = k_e x_e \left [n({\rm H}) + 2 n({\rm H}_2) \right ] {N(\htp) \over N(\Htwo)},
\end{align}

\noindent where $k_e = -1.3 \times 10^{-8} + 1.27 \times 10^{-6} \ T_e^{-0.48}$ is the rate coefficient for recombination of $\htp$ and $e^-$, $T_e \approx 70 \ {\rm K}$ is the electron temperature, $x_e \equiv n_e/(n({\rm H}) + 2 n({\rm H}_2))$ is the electron fraction, and $N(\Htwo)$ is the column density of $\Htwo$. The fractional abundance of electrons is well-approximated by that of ${\rm C}^+$, which has been measured in many lines-of-sight, with an average value of $x({\rm C}^+) \approx 1.5 \times 10^{-4}$. Upper limits on CR ionization rates in the direction of several clouds are listed in~\cite{IndrioloThesis} with values ranging from $\approx 3.3 \times 10^{-17} \ {\rm s}^{-1}$ to $\approx 2.8 \times 10^{-15} \ {\rm s}^{-1}$. 

\subsection{Observational details: molecular hydrogen ionization rate in dense MCs}

In dense clouds, ${\rm H}_3^+$ is no longer a good tracer for the CR ionization rate, and alternative tracers -- including HC$_3$N, HCO$^+$, DCO$^+$ (``D'' refers to deuterium), and CO -- must be used~\cite{Caselli1998}. The chemical network responsible for the production of these tracers is considerably more complex than that of ${\rm H}_3^+$ (see~\cite{LBH1996}). The free electron fraction and CR ionization rate in dense MCs can be determined by the the abundance ratios $R_D = [\text{DCO}^+]/[\text{HCO}^+]$ and $R_H = [\text{HCO}^+]/[\text{CO}]$~\cite{Butner1995}. Among the main uncertainties in the chemical model described in~\cite{LBH1996} is the effect of depletion of atomic C and O onto dust grains. The destruction of ${\rm H}_3^+$ depends sensitively on the abundance of gas phase C and O. Ref.~\cite{LBH1996} adopts $x({\rm O}) \approx x({\rm CO})$, which is later revised by~\cite{Caselli1998} to $x({\rm O}) = x({\rm CO})/f_D$, where $1/f_D$ is the fraction of O that remains in the gas phase after depletion. The free electron fraction and CR ionization rates can then be determined using

\begin{align}
    x_e &= {2.7 \times 10^{-8} \over R_D} - {1.2 \times 10^{-6} \over f_D}, \\
    \zeta^{{\rm H}_2} &= x_e n({\rm H}_2) R_H \left( 7.5 \times 10^{-4} x_e + {4.6 \times 10^{-10} \over f_D }\right).
\end{align}

Using these analytic expressions, Ref.~\cite{Caselli1998} determined the CR ionization rates in 24 dense cloud cores with large UV-optical attenuation. The clouds with the lowest CR ionization rates are L1551, L63, and L1262, however the latter suffers from large uncertainties in the gas depletion factor, $f_D$. The depletion factor in cloud L1551 was estimated using observations of the abundance ratio, $[{\rm HC}_3{\rm N}]/[{\rm CO}]$ and is consistent with $f_D \le 3$. Conservatively assuming $f_D = 3$, the ionization rate for L1551 is $ \log_{10}(\zeta^{{\rm H}_2} \cdot {\rm sec}) = -17.3 \pm 0.2$ and for L63 $ \log_{10}(\zeta^{{\rm H}_2} \cdot {\rm sec}) = -17.2 \pm 0.2$. In deriving constraints below, we focus exclusively on L1551 and adopt a value
\begin{equation}
    \log_{10}(\zeta^{{\rm H}_2} \cdot {\rm sec}) = -17.1 \implies \zeta^{{\rm H}_2} = 8\times 10^{-18}\, {\rm s}^{-1}.
\end{equation}

\section{Molecular Migdal Effect}
\label{sec:MolecularMigdal}

The Migdal effect generally refers to electronic excitation following nuclear scattering. In contrast to the Migdal effect in isolated atoms, which may be understood as a purely kinematic effect resulting from the small mismatch between the atomic center of mass and the position of the nucleus due to the nonzero electron mass~\cite{lovesey1982electron,Kahn:2021ttr}, the Migdal effect in molecules necessarily involves the interactions between nuclei and electrons. In that sense, the Migdal effect is a correction to the Born-Oppenheimer approximation, in which the electronic and nuclear motion is decoupled. In both atoms and molecules, the Migdal effect may lead to either bound transitions (excitation) or ionization, but in the molecular case, we must also include the effects of nuclear rotational and vibrational excitation, as well as dissociation, all of which may accompany the electronic transitions.

\label{sec:migdalrate}
\subsection{Bound transitions}
Here we briefly review the results of Ref.~\cite{Blanco:2022pkt}, which focused on bound Migdal transitions in molecules. While these do not contribute to the ionization rate in molecular clouds, they will inform our calculation of the Migdal ionization rate. The squared matrix element for a transition from the molecular ground state $|\Psi_0\rangle$ to a final state $|\Psi_\alpha \rangle$ in a homonuclear diatomic molecule, following a DM-nuclear scattering with momentum deposit $\vec{q}$, is
\be
\label{eq:PM}
P^{(\alpha)} = | \langle \Psi_\alpha | e^{i \vec{q} \cdot \vec{R}_1} + e^{i \vec{q} \cdot \vec{R}_2} | \Psi_0 \rangle |^2
\ee
where $\vec{R}_1$ and $\vec{R}_2$ are the nuclear coordinate operators. Ref.~\cite{Blanco:2022pkt} considered the case where $\Psi_\alpha$ is an excited electronic state of the bound molecule, in which case nuclear scattering can provoke electronic transitions through center-of-mass recoil (CMR) or non-adiabatic coupling (NAC). In general, the NAC Migdal effect is an order of magnitude larger than CMR, which is also the case for $\Htwo$~\cite{colognesi2005can}. As a result, we neglect the CMR matrix elements. 

The NAC matrix element in \htwo arises from corrections to the Born-Oppenheimer approximation, and is given by
\be
P^{(\alpha)}_{\rm NAC} = \sum_n |\mathcal M^{(\alpha)}_{{\rm NAC}, n}|^2 \equiv P^{(\alpha)}_{e,{\rm NAC}} \times P^{(\alpha)}_{N,{\rm NAC}},
\ee
with
\be
\label{eq:PeNAC}
P^{(\alpha)}_{e,{\rm NAC}} = \frac{q^2 \eta^2 |G_{\alpha 0}|^2}{M^2(\epsilon_\alpha - \epsilon_0)^2}
\ee
and
\be
P^{(\alpha)}_{N,{\rm NAC}} = 16  \sum_{n} | \langle \chi^{(\alpha)}_n | \sin (q \rho \eta/2)| \chi_0 \rangle|^2.
\ee
Here, $\eta = \hat{q} \cdot \hat{\rho}$ is the cosine of the angle between $\vec{q}$ and the molecular axis $\vec{\rho}$, $\epsilon_\alpha - \epsilon_0$ is the energy of the state $\Psi_\alpha$ above the ground state, $M$ is the proton mass, $\chi_0$ and $\chi_n^{(\alpha)}$ are the nuclear wavefunctions associated to the ground state and excited state respectively, and
\begin{align}
   G_{\alpha 0} &= \int\, \prod_i d^3\vec r_i \, \psi^*_{\alpha}(\vec r_i\text{;}\rho_{\alpha}) \left. \left(\vec \nabla_\rho \psi_0(\vec r_i\text{;}\rho)\right) \right |_{\rho = \rho_0} 
\end{align}
is the non-adiabatic coupling between $\Psi_\alpha$ and $\Psi_0$, with $\vec{r}_i$ the electronic coordinates and $\rho_0$ the equilibrium nuclear separation of the ground state.

Computing $G_{\alpha 0}$ requires a model for the \htwo wavefunctions. Ref.~\cite{colognesi2005can} used a MO-LCAO model, finding $G_{\alpha 0} \approx 0.05 \, \azero^{-1}$ for the lowest-energy NAC transition to the $E^{i 1}\Sigma_g^+ 2{\rm s}\sigma$ state, with $\epsilon_\alpha - \epsilon_0 = 12.409 \ {\rm eV}$. An approximate sum rule, which only requires knowledge of the ground-state wavefunction, may also be obtained for bound NAC transitions. The probability of transitioning to any bound NAC state above the ground state, inclusive of all nuclear transitions, is approximately
\be
P_{\alpha \neq 0, {\rm NAC}} \approx \frac{q^2}{8}G_0,
\label{eq:NACapprox}
\ee
where $G_0 \approx 1.69 \times 10^{-6} \, \azero^2$ for \htwo and is related to gradients of the ground-state wavefunction with respect to the nuclear separation.

\subsection{Ionized and dissociated transitions}

\begin{figure}
    \centering
    \hspace{-0.4cm}\includegraphics[width=0.5\textwidth]{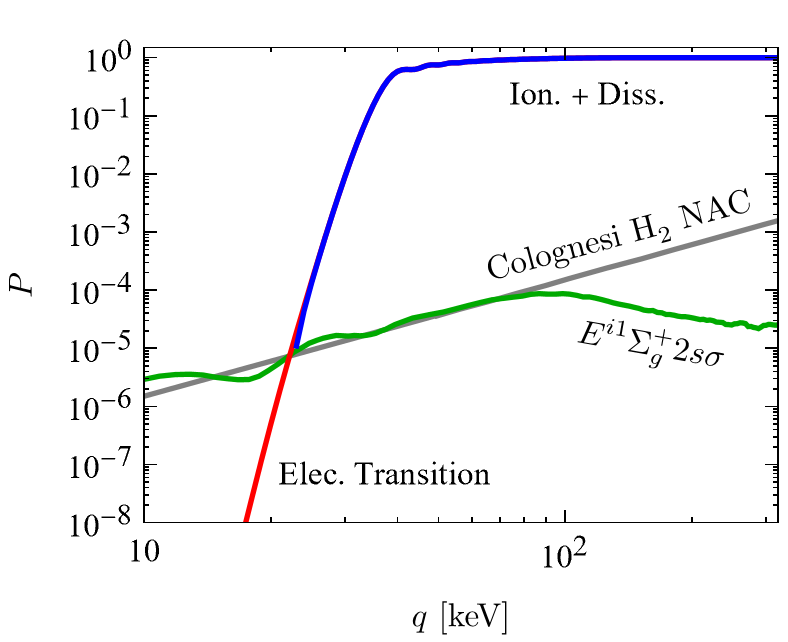}
    \caption{Transition probabilities in \htwo as a function of momentum transfer $q$ for the first NAC transition (green), all bound NAC transitions (gray) \cite{colognesi2005can}, inclusive ionization and dissociation transitions (blue), and all electronic transitions (red). Note that Eq.~(\ref{eq:NACapprox}) is an approximation to the inclusive excitation rate, so our exclusive calculation of the first NAC transition (green) exceeds the gray curve at small momenta. Below $\sim 20 \keV$, our inclusive probability of any electronic transition (red) drops below the exclusive excitation probability $P_{\rm{NAC}}$ (green), which indicates a breakdown of our approximations. As a result, we will only consider momentum transfers above 22 keV.}
    \label{fig:transition comparison}
\end{figure}

At sufficiently large momentum transfers $q$, the $\Htwo$ molecule is likely to be ionized or dissociated, with a probability approaching 1 as $q \to \infty$. Ionization frequently accompanies dissociation, and in the astrophysical context of CR scattering most relevant for this work, the former dominates the latter~\cite{ODonoghue_2022}. We also note that the rate of ionization dominates over dissociation by a factor of 2 to 10 in neutral-hydrogen collisions with $\Htwo$, which is the closest analog in astrophysical reactions to DM-$\Htwo$ collisions~\cite{padovani2018production,shaw2020cosmic,phelps1990cross}. As such, we will approximate the ionization rate as the total inclusive rate of ionization plus dissociation. To compute a total transition rate per molecule, we can compute an inclusive probability as follows:
\be
    P_{\textrm{ion.}}\approx 1-P_{\rm bound}- \sum_{n=0}^{n_{\textrm{dis.}}}\sum_{l=0}^{l_{\rm{dis.}}(n)}\sum_{m=-l}^{l}P_{e=e_0}(n,l,m).
\label{eqn:ion}
\ee
Here, $P_{\rm bound}$ is the probability of bound electronic excitation, which we can approximate as $P_{\alpha \neq 0, {\rm NAC}}$ as discussed above. The last term in Eq.~(\ref{eqn:ion}) is the probability of remaining in the ground electronic state $e = e_0$, summing over all rotovibrational transitions $n$ up to maximum vibrational level $n_{\textrm{dis.}}$ and maximum rotational level $l_{\rm{dis.}}$, where the cutoffs ensure that the total rotovibrational energy is less than the molecular dissociation energy. The angular momentum cutoff $l_{\rm{dis.}}$ is computed for a given $n$ using rigid rotor eigenstates as an approximation. We compute $P_{e=e_0}(n,l,m)$ in the Born-Oppenheimer approximation, taking the nuclear wavefunctions $\chi_n$ to be harmonic oscillator states $\chi_n$ times spherical harmonics, and $\Psi_0 = \psi_0(\vec{r}_i; \rho) \chi_0(\vec{\rho})$. For transitions where there is no electronic excitation, modeling of the electronic wavefunctions $\psi_0$ in the Born-Oppenheimer approximation is unnecessary because for a final state $\Psi'_0 = \psi_0 \chi_n Y^m_l$, we have
\be
\langle \Psi'_0 | e^{i \vec{q} \cdot \vec{R}_1} + e^{i \vec{q} \cdot \vec{R}_2} | \Psi_0 \rangle = \langle \psi_0 | \psi_0 \rangle \mel{\chi_n Y^m_l}{e^{i q \rho \eta/2}}{\chi_0 Y^0_0}
\label{eq:overlap}
\ee
and the electronic matrix element drops out because $\langle \psi_0 | \psi_0 \rangle = 1$ for any choice of $\psi_0$. Since the matrix element in Eq.~(\ref{eq:overlap}) is azimuthally symmetric and the ground state is isotropic, the transitions are restricted to $m=0$. The nuclear matrix elements may be evaluated analytically as described in Ref.~\cite{Blanco:2022pkt}, where explicit formulas may be found. Taking a ground-state vibrational energy of 0.55 eV and a dissociation energy of 4.5 eV, we find $n_{\rm dis.} = 7$ and $l_{\rm{dis}}$ ranging from 23 for $n=0$ to 6 for $n=7$. As discussed in~\cite{Blanco:2022pkt}, more accurate modeling of the nuclear wavefunctions is not expected to qualitatively change the result. We note, however, that the kinematic threshold for ionization depends strongly on the maximum $l$ in the sum, as accounting for increasingly large energy transitions requires increasingly large momentum transfer to leave the ground state. Our rigid rotor model could be further refined with more precise $\Htwo$ wavefunction modeling.

Fig.~\ref{fig:transition comparison} shows the transition probabilities as a function of $q$ for the exclusive lowest NAC transition averaged over $\eta$ (green), the inclusive NAC transition approximation from Ref.~\cite{colognesi2005can} (gray), all transitions from the ground electronic state including the bound transitions (red), and ionization plus dissociation (blue). At small $q$, the lowest NAC transition dominates the inclusive NAC probability, as expected. At $q \simeq 100 \ {\rm keV}$, other bound NAC transitions take over, but are still negligible compared to ionization and dissociation. This means that our estimate of the ionization rate (which involves only the nuclear wavefunctions) is relatively insensitive to the modeling of the electronic orbitals. As expected, the total inclusive excitation probability approaches 1 at large $q$, with the effect of the bound transitions negligible by about $q \simeq 30 \ {\rm keV}$. However, at around the same scale, the inclusive electronic transition probability (red) drops below the exclusive NAC excitation probability (green), indicating a breakdown of our approximations. Nonetheless, there is a large probability of ionization/dissociation at momenta above $\simeq 20 \ {\rm keV}$ which are accessible to sub-GeV DM. 

\subsection{Migdal ionization rate}


As mentioned in Sec. \ref{sec:CRionization}, sufficiently energetic DM impinging on a MC can ionize \htwo in the same way CRs do in (\ref{eqn:crchem}). The precise mechanism by which DM ionizes \htwo depends on the microphysical model leading to interactions between DM and the Standard Model. While previous work in Ref.~\cite{Blanco:2022mci} considered DM coupled only to electrons, here we consider a scenario in which DM couples to protons and ionizes $\Htwo$ through the Migdal effect as described above. In this case, the DM ionization rate per molecule is given by 
\be
\frac{R}{N_T}=\frac{\rho_{\chi} f_{\chi}}{m_{\chi}}\frac{ \sigma_p}{8\pi \mu_{\chi p}^2}\int_{q_{\rm min}} \frac{d^3q}{q}[P_{\rm{ion}}(q)]^2\eta(v_{\rm{min}})F_{\rm{DM}}^2(q),
\label{eq:rate}
\ee 
where $\rho_\chi = 0.3 \ {\rm GeV/cm}^3$ is the local DM density, $q_{\rm min} = 22 \ {\rm keV}$ is the minimum $q$ above which our approximations for computing $P_{\rm ion}(q)$ are valid, $f_{\chi}$ is the fractional DM abundance of the DM species contributing to Migdal-induced ionization, $\sigma_p$ is a reference DM-proton cross section, $\mu_{\chi p}$ is the DM-proton reduced mass, and $\eta$ is the mean inverse velocity
\be
\eta(v_{\rm{min}})=\int \frac{d^3v}{v}f_\chi(v)\Theta(v-v_{\rm{min}}).
\ee
For the DM velocity distribution in the frame of the cloud, $f_\chi(v)$, we take Standard Halo Model ansatz of a Maxwellian distribution in the galactic frame and boost by the velocity of the cloud:
\be
f_\chi(v)=\frac{1}{N_0}e^{-|\vec{v} + \vec{v}_{\rm cloud}|^2/v_0^2}\Theta(v_{\rm esc} - |\vec{v} + \vec{v}_{\rm cloud}|).
\ee
Here $v_0=230 \ {\rm km/s}$ is the velocity dispersion, $v_{\rm esc} = 600 \ {\rm km/s}$ is the Galactic escape velocity, $N_0$ is a normalization factor due to the truncation of the Maxwellian distribution by the escape velocity (see Ref.~\cite{Essig:2015cda} for an explicit expression), and the velocity of the L1551 MC with respect to the Galactic frame is~\cite{GalliMCVelocity,Blanco:2022mci}
\be
\vec{v}_{\rm cloud} = (-16 \hat{\rho} + 205 \hat{\phi} - 7 \hat{z}) \ {\rm km/s}
\ee
in Galactocentric coordinates. The minimum velocity to ionize hydrogen, $v_{\rm min}$ in Eq.~(\ref{eq:rate}), is given by
\be
v_{\rm{min}}=\frac{q}{2\mu_{\chi T}}+\frac{E_{\rm{ion}}}{q},
\ee
where $\mu_{\chi_T} = \frac{m_\chi m_{\Htwo}}{m_\chi + m_{\Htwo}}$ is the DM-target reduced mass, and $E_{\rm{ion}} = 15.4 \ {\rm eV}$ is the ionization energy of $\Htwo$.

Finally, $F_{\rm{DM}}^2(q)$ in Eq.~(\ref{eq:rate}) is the DM form factor which parameterizes the momentum dependence of the DM-proton interaction. We take two representative models: a heavy mediator which yields a momentum-independent contact interaction $F_{\rm DM} = 1$, and a massless mediator which yields $F_{\rm DM} \propto 1/q^2$. In the latter case, a reference cross section must be defined at an arbitrary momentum scale $q_0$. A representative model is one in which DM has millicharge $\epsilon e$ under ordinary electromagnetism, in which case the mediator is the photon and
\be
\label{eq:millicharge}
\frac{\sigma_p}{\mu_{\chi p}^2}  = \frac{\epsilon^2 e^4}{\pi q_0^4}, \qquad F_{\rm DM} = \frac{q_0^2}{q^2}.
\ee

To avoid referring to the unphysical fiducial momentum $q_0$, and to eliminate the need to compare differing conventions in the literature for light-mediator cross sections with protons, we will present all bounds for the light mediator model in terms of the millicharge parameter $\epsilon$.

The observed ionization rate, $\zeta_\text{obs}^{\text{H}_2}$ should be equal to the sum of $\zeta_i^{\text{H}_2}$ over all species. We set a conservative upper bound on the DM-proton interaction cross section by requiring that $\zeta_\text{DM}^{\text{H}_2} < \zeta_\text{obs}^{\text{H}_2}$. The larger incident flux and lower cross section means that more DM particles than CRs can penetrate deep into a cloud. In the following sections, we directly compute $\zeta_\text{DM}^{\text{H}_2}$ and use observations of $\zeta_\text{obs}^{\text{H}_2}$ to place constraints on the interaction strength between DM and protons.

\section{Results}
\label{sec:results}

\begin{figure}
    \centering
    \includegraphics[width=0.48\textwidth]{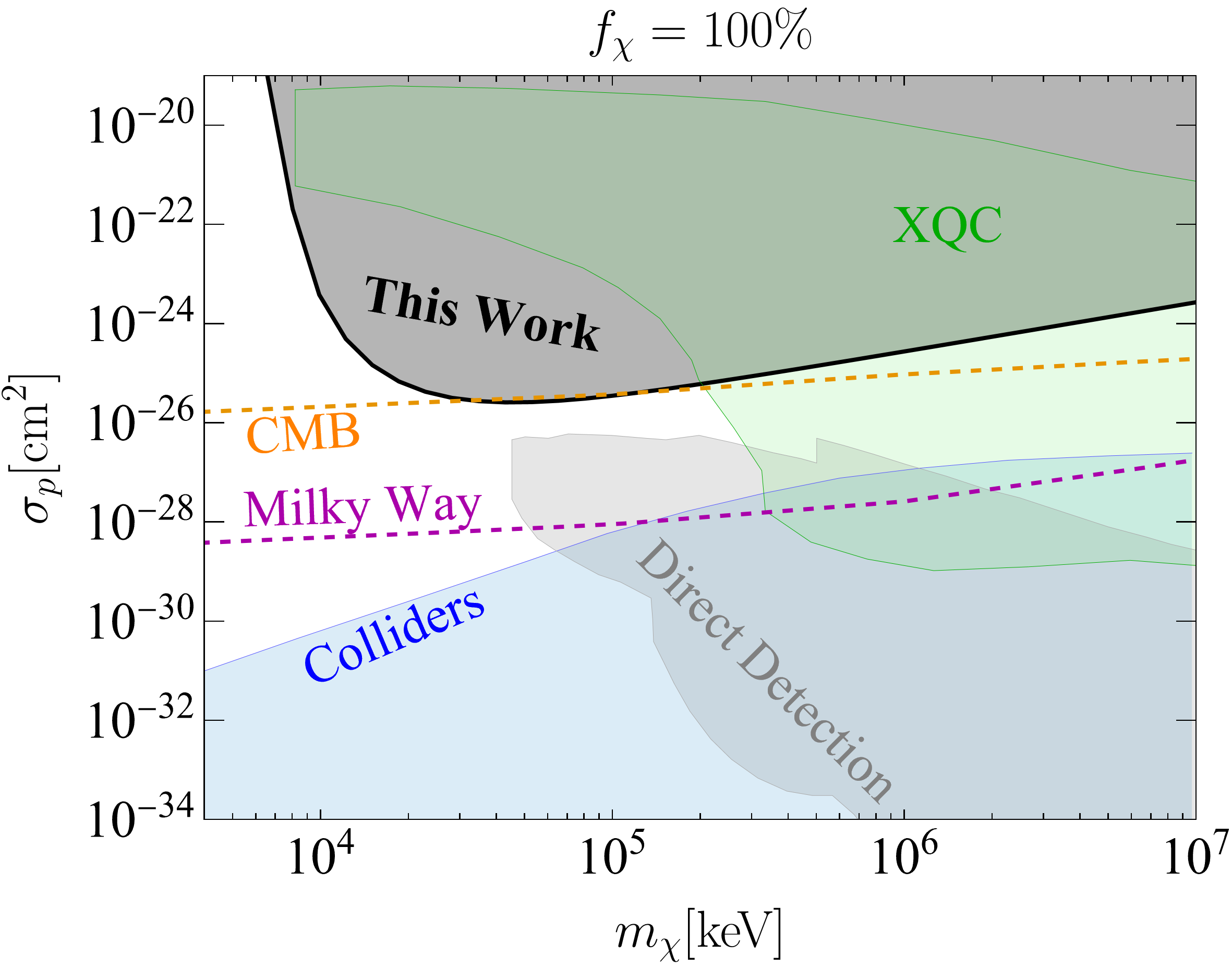}
    \caption{Constraints from Migdal scattering in the L1551 cloud, in the heavy mediator model with $F_{\rm DM}=1$ (dark shaded gray). Constraints are shown compared to existing bounds from direct detection (ligh gray) \cite{Emken_2018}, CMB (dashed orange) \cite{Boddy_2022}, the X-Ray Quantum Calorimeter (XQC) experiment~\cite{Li:2022idr} (solid green), collider bounds \cite{Daci_2015} (solid blue), and Milky Way satellite galaxy bounds from the Dark Energy Survey and Pan-STARRS1 \cite{Milky_Way} (dashed purple). Dashed bounds correspond to those which are severely weakened with a sufficiently small DM subfraction $f_\chi$.}
    \label{fig:sensitivity}
\end{figure}

To calculate the sensitivity to DM-proton scattering in the L1551 MC, we use a detection threshold corresponding to a per-molecule ionization rate of $\zetacr=8\times 10^{-18}\ {\rm s}^{-1}$, in line with Sec.~\ref{sec:CRionization}. We first set $f_{\chi}=1$. The results for a heavy mediator ($F_{\rm DM} = 1$) are shown in Fig.~\ref{fig:sensitivity}. While Migdal ionization covers a large region of strongly-interacting parameter space unprobed by collider or direct detection experiments, constraints on DM-baryon scattering from the CMB \cite{Boddy:2018wzy} and Milky Way satellites \cite{Milky_Way} surpass Migdal ionization for a wide range of masses. However, as previously mentioned, the CMB bounds vanish for $f_{\chi}\lesssim 0.4\%$ \cite{Boddy:2018wzy}, and for similar reasons the Milky Way satellite bounds should vanish at sufficiently small $f_\chi$ on the order of $0.1\%$, though the precise value requires detailed numerical simulations. Thus, for $f_\chi$ below that critical value, Migdal ionization may provide the leading constraint on spin-independent DM-proton scattering, because the Migdal ionization rate scales linearly with the DM subfraction down to arbitrarily small values of $f_\chi$ so long as this DM component still has a roughly Maxwellian velocity distribution. 

\begin{figure}
    \centering
    \includegraphics[width=0.48\textwidth]{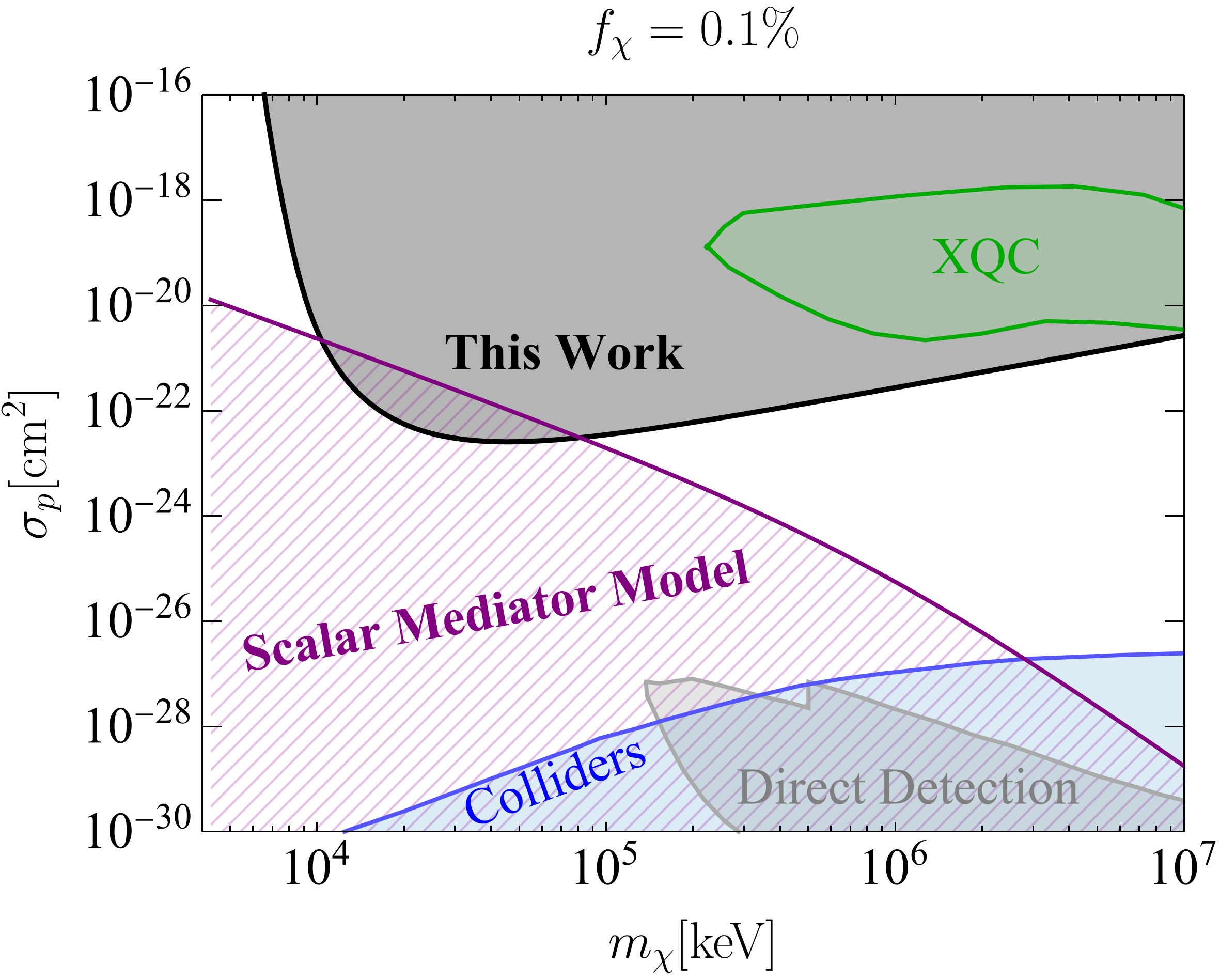}
    \caption{Constraints on a DM subfraction of $f_\chi = 0.1$\%, scattering through a heavy mediator (dark shaded gray). Other bounds which scale with $f_\chi$ have been adjusted accordingly; we do not show the CMB and Milky Way bounds which will disappear for small $f_\chi$ on the order of 0.1\%. The purple hashed region shows the allowed parameter space for the scalar mediator model in Eq.~(\ref{eq:HM}). The MC constraints we derive in this work rule out a portion of this parameter space for masses around 10--100 MeV.}
    \label{fig:subfraction}
\end{figure}

Fig.~\ref{fig:subfraction} shows our constraints for a subfraction $f_\chi = 0.1\%$. We see that MCs can probe considerable areas of strongly coupled parameter space, particularly for DM masses from 10--100 MeV. At higher masses, direct detection constraints from the XQC satellite~\cite{Li:2022idr} provide the leading constraints, though the observed event rate at the detector is low enough that the bounds disappear entirely for $f_\chi \lesssim 10^{-6}$. Given the rather large cross sections which are accessible to MCs, it is important to identify a benchmark particle physics model which may generate such large cross sections while still remaining consistent with astrophysical, cosmological, and laboratory constraints. As an example of such a model, consider a real scalar DM field $\chi$ that couples to nucleons $n$, through a scalar mediator $\phi$ which couples to gluons at high energies. The low-energy Lagrangian is
\be
    \LL \supset -\frac{1}{2}m_{\chi}^2 \chi^2 - \frac{1}{2}m_{\phi}^2 \phi^2 -\frac{1}{2}y_{\chi}m_{\chi}\phi \chi^2 -y_n \phi\bar{n}n,
    \label{eq:HM}
\ee
where $y_\chi$ and $y_n$ are dimensionless couplings. This model, which was studied in detail in Ref.~\cite{Knapen:2017xzo}, yields a non-relativistic DM-proton cross section
\be
    \sigma_{\chi p}\equiv \frac{y_p^2y_{\chi}^2}{4\pi}\frac{\mu^2_{\chi p}}{\left(m_{\phi}^2+v_{\chi}^2m_{\chi}^2 \right)^2},
    \label{eq:heavycrosssec}
\ee
where $y_\chi, y_n \lesssim 4\pi$ for the theory to remain perturbative. The mediator is ``heavy'' so long as its mass is comparable to the DM momentum scale, $m_\phi \gtrsim v_{\chi} m_{\chi}$, where $v_\chi \sim 10^{-3}$ is the typical DM velocity. The leading constraints on this model, for $m_\chi$ in the mass range which could be probed by MCs, come from rare kaon decays $K \to \pi \phi$, which would appear as an excess in $K \to \pi \nu \bar{\nu}$~\cite{NA62:2021zjw}. The resulting limit is $y_n \lesssim 7.3 \times 10^{-6}$. Setting $y_n$ to the maximum value allowed by kaon decays, along with $m_\phi \geq 10^{-3} m_\chi$ and $y_\chi \leq 4\pi$, thus bounds the regime of validity for this model.\footnote{Note that with a subfraction $f_\chi = 0.1\%$, DM self-interaction constraints are not relevant.} The purple region in Fig.~\ref{fig:subfraction} shows the allowed region of parameter space for this scalar DM model; our MC constraints can still rule out portions of this parameter space for $m_\chi$ in the 10--100 MeV range.

For a light mediator, we take the millicharged DM model given in Eq.~(\ref{eq:millicharge}) and place constraints on the DM millicharge $\epsilon$. While CMB bounds once again surpass our MC bounds for $f_{\chi}=1$, these bounds vanish for $f_{\chi}\lesssim 0.4\%$ \cite{Boddy:2018wzy}, while our sensitivity would linearly decrease as before. Our Migdal bounds are slightly weaker by an $\mathcal{O}(1)$ factor compared to the bounds from DM-electron scattering in MCs from Ref.~\cite{Blanco:2022mci}. This difference is most apparent at the smallest DM masses, because the breakdown of our approximations for the inclusive excitation rate imposes a momentum cutoff $q > 22 \ {\rm keV}$, which effectively sets a 
DM threshold of $m_\chi = 5.6 \ {\rm MeV}$ which is above the kinematic ionization threshold of $\sim 4 \ {\rm MeV}$. The dominance of electron scattering bounds compared to Migdal ionization for a light mediator is consistent with the results of Ref.~\cite{Baxter:2019pnz} in the direct detection context. Since the electron scattering bounds and the Migdal bounds both scale linearly with $f_\chi$, for small DM subfractions the electron scattering bounds will continue to dominate, and Ref.~\cite{Blanco:2022mci} thus gives the strongest constraints on this model.

\begin{figure}
    \centering
    \includegraphics[width=0.48\textwidth]{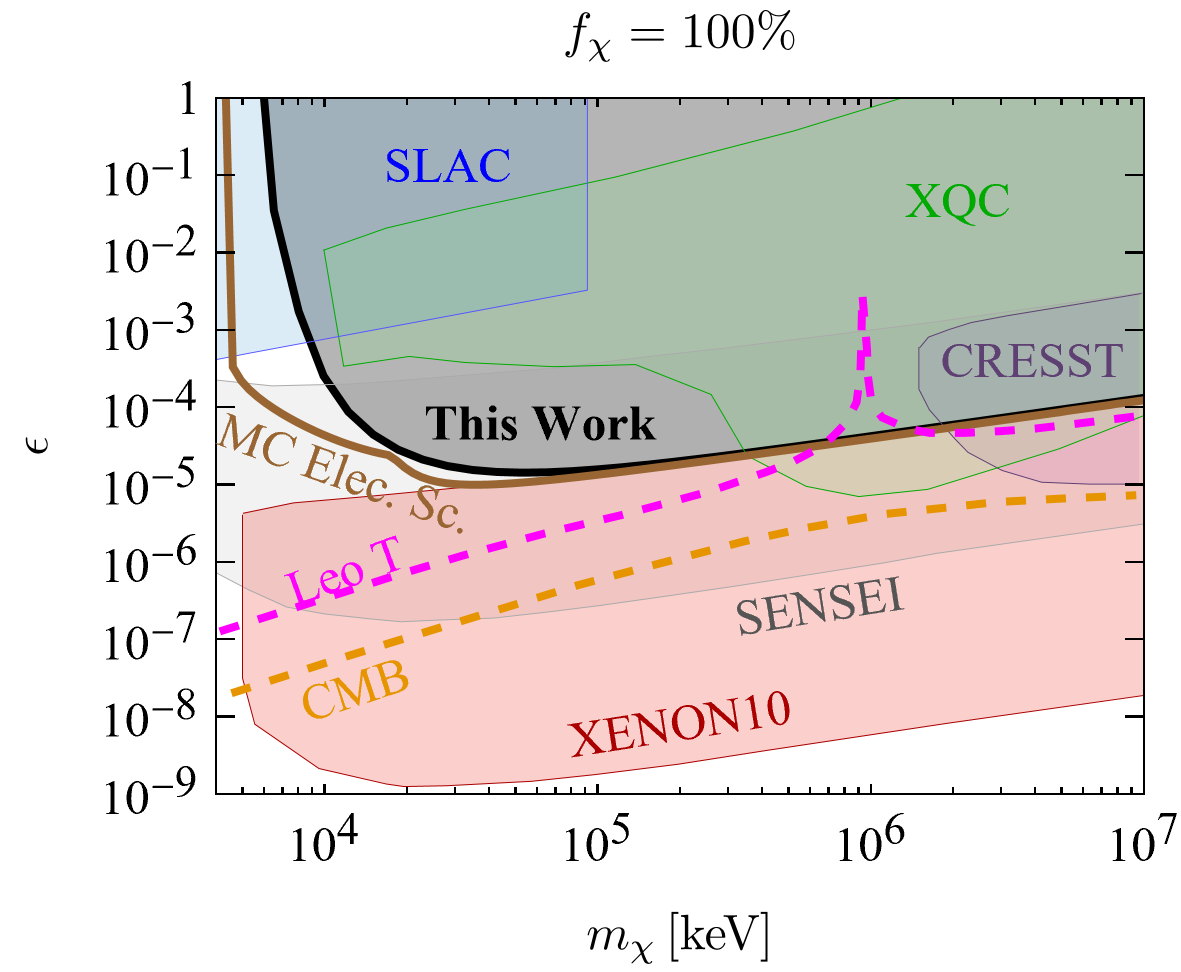}
    \caption{Constraints from Migdal scattering in the L1551 cloud, in the millicharged DM model (dark shaded gray). Bounds are shown compared to those from existing experiments: direct detection (CRESST \cite{Angloher_2017}, SENSEI \cite{Crisler_2018}, and XENON10 \cite{Essig_2012,Essig_2017})), astrophysical observations (gas-rich galaxy Leo T \cite{Wadekar:2019mpc}, XQC \cite{Li:2022idr}), and the SLAC millicharge search \cite{Prinz:1998ua}. CMB bounds \cite{Creque_Sarbinowski_2019,Mu_oz_2018}, which disappear for a DM subfraction $f_{\chi}\lesssim 0.4\%$, are shown in dashed orange. $\rm{H}_2$ MC bounds for electronic scattering, taken from \cite{Blanco:2022mci} and adjusted for our chosen MC ionization rate, are shown in brown for comparison.}
    \label{fig:sensitivityMC}
\end{figure}

\section{Conclusion}
\label{sec:discussion}

In this work, we have presented constraints on the DM-proton coupling for sub-GeV-scale DM using measurements of ionization rates in MCs. Our constraints complement similar constraints on DM-electron coupling in leptophilic DM models~\cite{Blanco:2022mci}. In our scenario, DM ionization of molecular hydrogen proceeds through the molecular Migdal effect~\cite{Blanco:2022pkt}, and is, to our knowledge, the first astrophysical application of this effect. Our analysis probes a regime of strongly-coupled DM that is inaccessible to terrestrial DM experiments. At large fractional abundance of this strongly-coupled DM, constraints from MC ionization complement but do not exceed constraints from CMB anisotropy measurements and observations of the Milky Way satellite population. However, if this strongly-coupled component makes up a small enough subfraction of DM, constraints from CMB anisotropies and Milky Way satellites disappear as this component of DM becomes degenerate with baryons. In contrast, constraints from MC ionization scale linearly with $f_\chi$ for $f_\chi \ll 1$ and would provide the strongest constraints on this sub-population. This work motivates the search for additional settings in which DM-induced ionization may have observable effects.

\section{Acknowledgements}
We thank Alex Drlica-Wanger, Ben Lillard, Ethan Nadler, Jes\'{u}s P\'{e}rez-R\'{i}os, and Benjamin Safdi for helpful discussions. A.P. acknowledges support from the Princeton Center for Theoretical Science postdoctoral fellowship. The work of C.B.~was supported in part by NASA through the NASA Hubble Fellowship Program grant HST-HF2-51451.001-A awarded by the Space Telescope Science Institute, which is operated by the Association of Universities for Research in Astronomy, Inc., for NASA, under contract NAS5-26555. The work of I.H. and Y.K. was supported in part by DOE grant DE-SC0015655. Part of this work was done the Aspen Center for Physics, which is supported by National Science Foundation grant PHY-1607611.

\bibliography{main.bib}
 
\end{document}